\documentclass[aps,pre,reprint,showpacs]{revtex4-1}
\usepackage{graphicx}
\usepackage{hyperref}
\hypersetup{
	colorlinks=true,
	urlcolor=blue
}
\usepackage{amsmath}

\begin{document}

\title{Self-propelled chimeras}

\author{Nikita Kruk\(^{1}\), Yuri Maistrenko\(^{1,2}\), and Heinz Koeppl\(^{1,}\)}
\email{Corresponding author. \\ heinz.koeppl@bcs.tu-darmstadt.de}
\affiliation{\(^{1}\)Department of Electrical Engineering and Information Technology, Technische Universit\"{a}t Darmstadt, Rundeturmstrasse 12, 64283, Darmstadt, Germany}
\affiliation{\(^{2}\)Institute of Mathematics and Center for Medical and Biotechnical Research, National Academy of Sciences of Ukraine, Tereshchenkivska St. 3, 01601, Kyiv, Ukraine}

\date{\today}

\begin{abstract}
  The synchronization of self-propelled particles (SPPs) is a fascinating instance of emergent behavior in living and man-made systems, such as colonies of bacteria, flocks of birds, robot ensembles, and many others. The recent discovery of chimera states in coupled oscillators opens up new perspectives and indicates that other emergent behaviors may exist for SPPs. Indeed, for a minimal extension of the classical Vicsek model we show the existence of chimera states for SPPs, i.e.,  one group of particles synchronizes while others wander around chaotically.  Compared to chimeras in coupled oscillators where the site position is fixed, SPPs give rise to new distinctive forms of chimeric behavior.  We emphasize that the found behavior is directly implied by the structure of the deterministic equation of motion and is not caused by exogenous stochastic excitation. In the scaling limit of infinitely many particles, we show that the chimeric state persists. Our findings provide the starting point for the search or elicitation of chimeric states in real world SPP systems.
\end{abstract}


\maketitle

\section{Introduction}

Collective behavior of large scale ensembles of agents is ubiquitous in nature. It is often characterized by the emergence of regular spatio-temporal dynamics from a disordered state without any central coordination. Examples include colonies of bacteria \cite{goldstein2012,peng2016}, schools of fish \cite{couzin2002}, flocks of birds \cite{cavagna2014}, groups of people \cite{silverberg2013,durrett2012}, and many more \cite{vicsek:phys_rep}. A model to analyze such dynamics is a system of SPPs that align the direction of motion to the average heading in their neighborhood. Under such an update rule \cite{vicsek:prl,ginelli2015}, the system either converges to complete alignment or remains in a disordered state. However, many of the experimentally observed collective dynamics do not fall into these two categories. More specifically, regular, coherent and irregular, disordered dynamics are seen to be present simultaneously.

Such coexistence of two disparate dynamical regimes is indicated, for instance, in the milling of a small group of fish within a large school \cite{chate2014} or in the vortexing of microtubuli within a large collection of meandering microtubuli \cite{sumino2012} or in the rotation of energized ferromagnetic colloids \cite{kaiser2017,geyer:2018}. Related complex dynamics, such as rotating chains or movings bands \cite{chate2008} can only be reproduced in SPP models in the presence of a strong stochastic driving term. A genuine coexistence of dynamical regimes in the absence of any stochastic forcing has recently been found in networks of non-locally coupled oscillators. In such a {\it chimera state}, groups of oscillators are synchronized while other oscillators undergo chaotic dynamics. This regime was first observed in the Kuramoto-Sakaguchi model \cite{kuramoto2002,abrams2004} of coupled phase oscillators that can be derived from the complex Ginzburg-Landau equation \cite{kuramoto1984}. The existence of chimera states has also been experimentally confirmed in optical \cite{hagerstrom2012,larger2015}, chemical \cite{tinsley2012,schmidt2014} and mechanical \cite{martens2013,kapitaniak2014} systems.

The paper is organized as follows. In Section II, we introduce our model for the spatial dynamics of self-propelled particles, and define the novel chimera states that are generated by that model. In Section III, we describe how such chimera states arise and what their structure is. Based on that knowledge, we discuss summary statistics in Section IV that can be applied in order to characterize the chimera states effectively. Afterwards, we show in Section V that the chimeras can be found in other setups too, especially in case of stochastic excitation. The introduced modifications allow us to consider the situation of infinitely many particles. We use that fact in Section VI in order to derive a continuum representation of the model. In Section VII, we summarize our results.

\section{Model}

We consider the following minimal extension of the Vicsek model \cite{vicsek:prl} in continuous time and in polar coordinates \cite{chepizhko:2010} for the velocity component, where $N$ particles move with a constant speed in a unit square domain with periodic boundary conditions, according to the equations of motion given by
\begin{equation}
\label{eq:spc_sys}
\dot{r}_{i} \,=\, v(\varphi_{i}),\quad
\dot{\varphi}_{i} \,=\, \frac{\sigma}{\left\vert B_{\rho}^{i} \right\vert} \sum_{j \in B_{\rho}^{i}}  \sin(\varphi_{j} - 
\varphi_{i} - \alpha)
\end{equation}
with ${r}_{i} = (x_{i}, y_{i})$ and ${v}(\varphi_i) \,=\, (\cos\varphi_{i}, \sin\varphi_{i})$, and the particles are assumed to have unit mass and unit speed, without loss of generality. Each particle $i$ interacts with all of its neighbors $j$ within a finite interaction range $\rho$, i.e., with all particles falling in the disk
\begin{equation*}
	B_{\rho}^{i} := \{j \mid (x_{i} - x_{j}) ^ 2 + (y_{i} - y_{j}) ^ 2 \leq \rho^2\}.
\end{equation*}
The alignment is controlled by the coupling coefficient $\sigma$ and the size of neighborhood $\left\vert B_\rho^i \right\vert$. We can consider equation \eqref{eq:spc_sys} as a generalization of the Kuramoto model in the sense that oscillators are augmented to be motile \cite{leonard2012, okeeffe2017}. Following this idea, we introduce the additional phase lag parameter $\alpha$ that originally allowed to observe chimera states in the Kuramoto-Sakaguchi model. In the context of SPPs, this parameter eventually induces a circular motion for the aligned group of particles. When $\alpha = 0$, the dynamics reduces to the Vicsek model in polar coordinates \cite{degond_motsch, chepizhko:phys_a}. As mentioned earlier, the uniqueness of our model is that it admits the coexistence of aligned and non-aligned collectives of particles. We call such a behavior a {\it self-propelled chimera state}.

\begin{figure}[!t]
	\centering
	\includegraphics[width=0.5\textwidth]{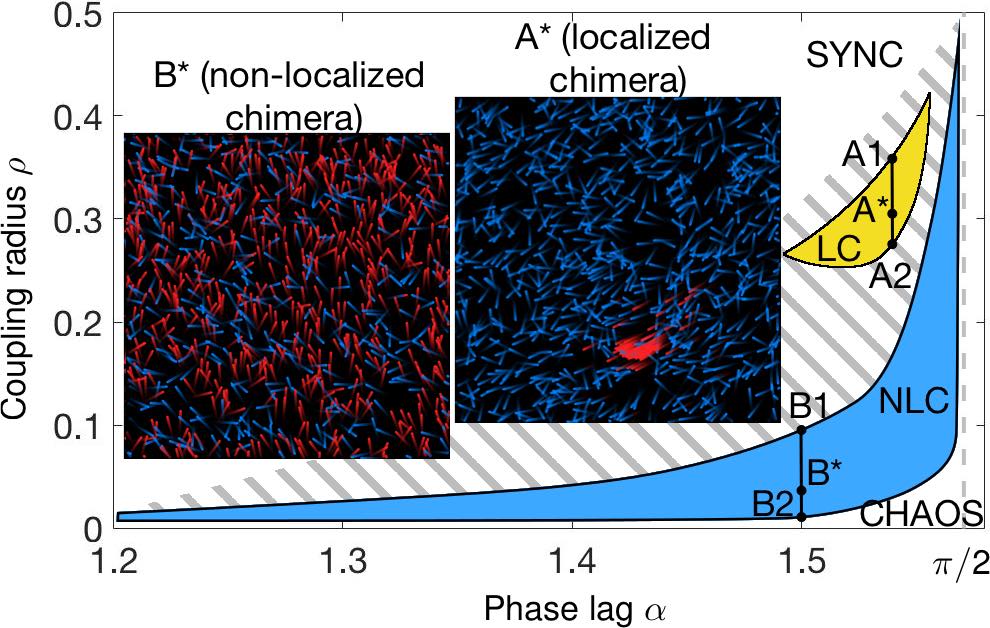}
	\caption
	{
		\label{fig:phase_diagram}
		Phase diagram for model \eqref{eq:spc_sys} in the $(\alpha,\rho)$ parameter plane. Localized (LC) and non-localized (NLC) chimeras exist in yellow (light gray) and blue (dark gray) regions, respectively. Snapshots demonstrate typical chimera regimes, i.e., A*=$(1.54,0.3)$, B*=$(1.5,0.03)$, respectively. Particles are colored with respect to averaged phase velocity $\left<\dot{\varphi}\right>$ (cf. Fig.~\ref{fig:space_time_plots}(a) and (c); averaging time is $t=5$ time units; see videos S1 and S2 under \cite{supp_mat,bcs_youtube_channel}) subject to binary thresholding. Intermittent behavior is characteristic for the neighboring region (oblique hatching). The region to the left leads to complete phase synchronization (SYNC) and the region to the right leads to full disorder (CHAOS). The lines A1-A2 and B1-B2 are used in Fig.~\ref{fig:summary_statistics} for order parameter description.
		Other parameters are $\sigma=1.0, N=1000$.
	}
\end{figure}

It has been shown that in the presence of noise the standard Vicsek model exhibits the formation of localized, traveling, high-density, and high-order structures, such as bands and sheets, or even blobs due to hydrodynamic long range interactions, but at sufficiently large noise amplitudes  \cite{chate2008,nagai2015}. We emphasize that our situation is different: coherent localized structures due to the model \eqref{eq:spc_sys} arise solely because of internal non-linear interactions imposed by non-local coupling in the complete absence of noise. It should be mentioned that interesting patterns such as traffic jams, gliders, and static bands can be found for a simple swarming model with ferromagnetic alignment mechanism and volume exclusion \cite{peruani2011}; however, these patterns do not constitute chimera states since such a model does not introduce phase synchronization. Moreover, if the alignment and anti-alignment are controlled depending on the range of interaction, coherent structures such as periodic vortex arrays may be produced \cite{grossmann2014}.

\section{Behavior}

Results of direct numerical simulation in the two-parameter plane of coupling radius $\rho$ and phase lag $\alpha$ are presented in the phase diagram in Fig.~\ref{fig:phase_diagram} obtained with the help of the continuation method (see the details of its implementation in Appendix A). The diagram reveals the existence of different chimera states in a considerable domain at intermediate radii $\rho$ and at phase lags $\alpha$ close to $\pi/2$. For smaller $\alpha$ or larger $\rho$ (the top left region), complete phase synchronization occurs, which is an analog to the standard Vicsek model.
On the contrary, for $\alpha$ close to $\pi/2$ and very small $\rho$, as well as for $\alpha \geq \pi/2$ the behavior of the system is chaotic (the region on the right).

\begin{figure*}[!t]
	\centering
	\includegraphics[width=1\textwidth]{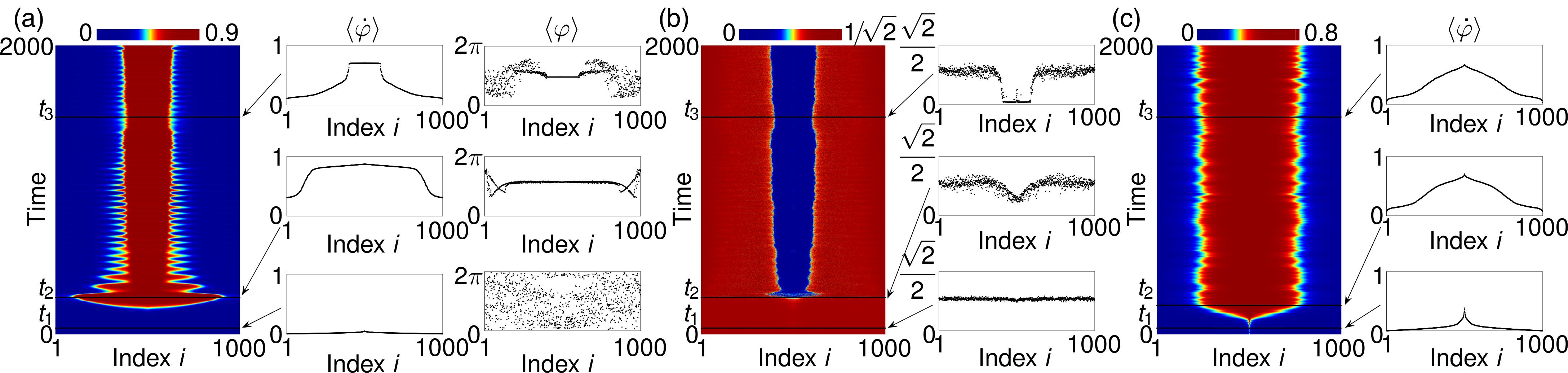}
	\caption
	{
		\label{fig:space_time_plots} Emergence of chimeras.
		(a) Space-time plot for LC, based on the averaged phase velocity. Along with it, profiles of averaged phase velocity $\left< \dot{\varphi} \right>$ and averaged phase $\left< \varphi \right>$ are provided for time points from characteristic periods of formation. 
		(b) Space-time plot for LC, based on the averaged value of distance from a particle position to a point of maximum density. Along with it, the corresponding profiles are provided at the same times as in (a). 
		(c) Space-time plot for NLC in the same manner as in (a).
		Particles are reordered identically at every time step with respect to the value of averaged angular velocity from (a) or (c),
		(b) is ordered according to (a).
		Temporal averaging is $t = 10$. Time points for (a) and (b) are $t_1 = 50$, 
		$t_2 = 255$, and $t_3 = 1500$; time points for (c) are $t_1 = 50$, $t_2 = 210$, 
		and $t_3 = 1500$. Parameters: 
		(a),(b) $\sigma = 1.0, \rho = 0.3, \alpha = 1.54$; 
		(c) $\sigma = 1.0, \rho = 0.03, \alpha = 1.5$.
	}
\end{figure*}

We observe two types of chimera states (see videos S1 and S2 under \cite{supp_mat,bcs_youtube_channel} for their temporal dynamics). Both types are classified as chimeras since they possess the partial synchronization property with respect to the direction of motion $\varphi$. This property is similar to phase synchronization of the Kuramoto-Sakaguchi model. However, the addition of the spatial equations reveals new possibilities for the group behavior. The first chimera type is characterized by the formation of a peculiar coherent and localized group. We call it a localized chimera (see inset \textit{A*} in Fig.~\ref{fig:phase_diagram}, and Fig.~\ref{fig:space_time_plots}(a),(b)) and refer to it throughout the paper as LC. Notably, this regime exists for a parameter region of only an intermediate interaction range $\rho$ (Fig.~\ref{fig:phase_diagram}, yellow (light gray) region). The second type of chimeras is characterized by partial phase synchronization but without any spatial localization. We call it a non-localized chimera (see inset \textit{B*} in Fig.~\ref{fig:phase_diagram}, and Fig.~\ref{fig:space_time_plots}(c)) and refer to it throughout the paper as NLC. It can be obtained for a distinct parameter domain of relatively larger size (blue (dark gray) region). In addition to the above two, the oblique hatched region corresponds to an intermittent system behavior between other states. Interestingly, at the center of this domain, multi-headed chimeras, which comprise two or more separate coherent groups, can be found. However, such multi-clustered structures are unstable and always disintegrate. Thus, we do not focus on them further.

In all the simulations, the initial positions and directions of particles were drawn from uniform distribution if not mentioned otherwise. First, we describe the evolution scenario for the LC case. In the initial stage (Fig.~\ref{fig:space_time_plots}(a),(b) at $t_1$), while the system tries to synchronize, 
none of the particles show any considerable increase in phase velocity (a flat $\left< \dot{\varphi} \right>$ profile), there is no visible group having the same traveling direction (a scattered $\left< \varphi \right>$ profile), and there is no a priori benchmark point to calculate the localization measure (a profile in Fig.~\ref{fig:space_time_plots}(b)). Meanwhile, particles gradually polarize and when the polarization is sufficiently large, the particles try to form a huge disk-shaped group, whose radius corresponds to the radius of interaction $\rho$. That group consists of the majority of particles. 
The profiles in Fig.~\ref{fig:space_time_plots}(a) at $t_2$ reveal its emergence. All particles in that group are directed similarly and are synchronized (flattened regions in the middle of the profiles). It appears that the group can not be maintained for a long time and many particles leave it until a smaller highly dense spot remains. That spot is stable and it is well distinguished from the other particles by the plateaus in each profile of Fig.~\ref{fig:space_time_plots}(a),(b) at $t_3$. 

The evolution of NLC is qualitatively similar till the so-called point of maximum synchronization (see the definition in the next section). After that point (Fig.~\ref{fig:space_time_plots}(c) at $t_2$), the dense disk-shaped group does not emerge. The system stalls in such a situation and the synchronized but scattered group remains.

\begin{figure}[!t]
	\centering
	\includegraphics[width=0.5\textwidth]{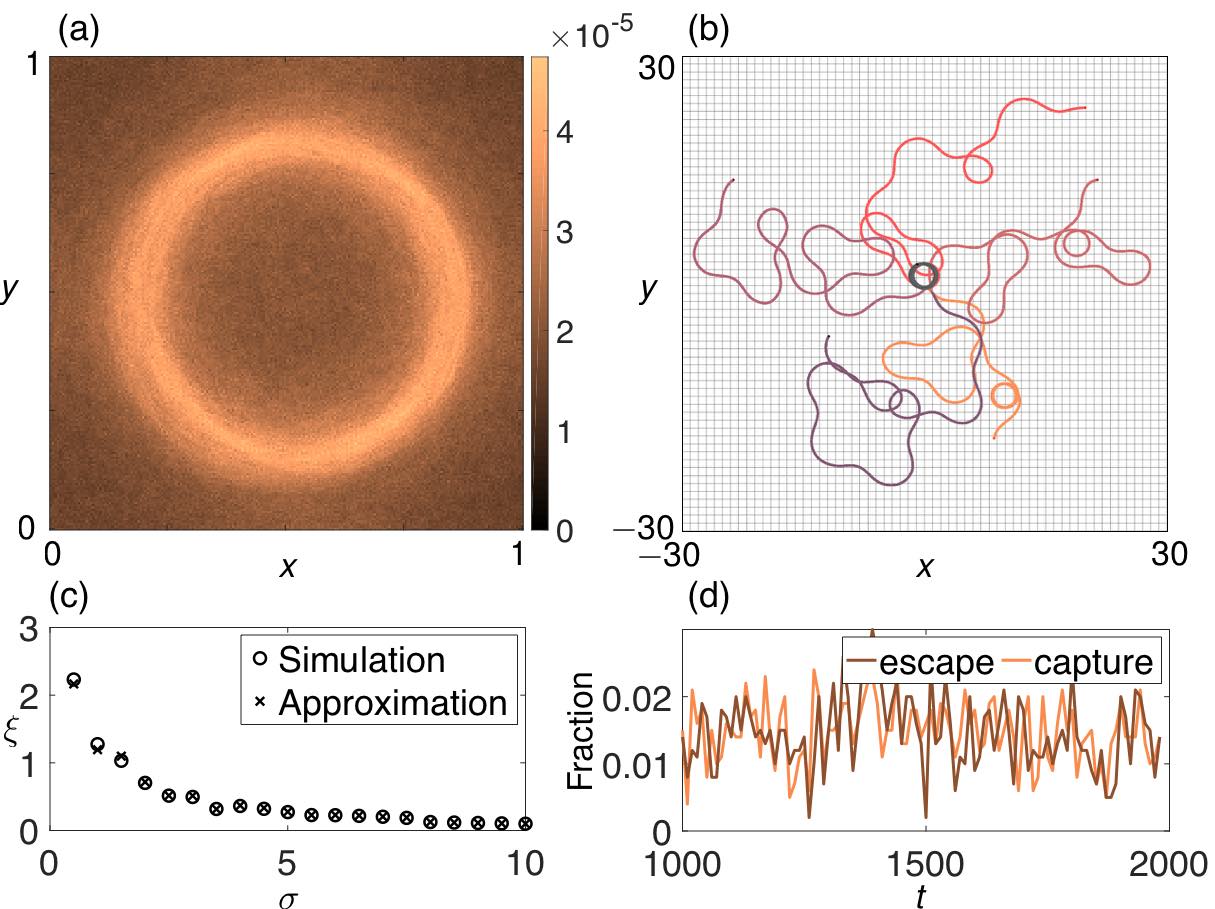}
	\caption
	{
		\label{fig:localization_traits}
		Localization traits of LC. (a) Traces of all particles over time ($t = 100$). Light circular trajectories indicate the motion of particles from a synchronized group whilst the dark background of considerable non-zero mass indicates the uniform spreading of the rest of chaotic particles. The color code corresponds to the number of particles at a particular position, normalized by the number of particles $N$ and by the averaging time. Coupling strength is $\sigma = 4$ in order to keep the whole circular trajectory inside a simulation box. 
		(b) Image of trajectories for one particle from a localized cluster 
		(black circle in the middle) and for five different particles from the chaotic cluster (here $\sigma = 1$) for unrolled periodic boundary conditions. 
		The tracking time is $t = 100$. 
		(c) Radius of rotation $\xi$ for particles from a localized group as a function of the coupling strength $\sigma$. 
		(d) Number of particles, which leave and join the synchronized cluster, per cluster cycle ($\approx 10$ time units) per population size.
		Other parameters are $\rho = 0.3, \alpha = 1.54, N = 1000$.
	}
\end{figure}

In the case of the scenario with localization, when the dense spot of synchronized particles appears, particles from that spot follow a quasi-circular trajectory while others fill the rest of the space uniformly (cf. Fig.~\ref{fig:localization_traits}(a),(b)). In such a setup, the phase dynamics in that synchronized group can be approximated as a combination of coherent and incoherent terms. A coherent term is imposed by all the particles which constitute the group. An incoherent term is imposed by all other desynchronized particles. Thus, we have
$
\dot{\varphi}_{i} = - \sigma \gamma_i \sin\alpha  + \frac{\sigma}{\left| B_{\rho}^{i} \right|} \sum_{j \in B_{\rho}^{i} \setminus N_{c}} \sin(\varphi_{j} - \varphi_{i} - \alpha),
$
where 
$\gamma_i = \left\vert N_c^i \right\vert / \left\vert B_\rho^i \right\vert$ is the fraction of coherent particles $N_c^i$ in the neighborhood $B_\rho^i$ of the particle $i$. Since every synchronized particle moves approximately along a circular trajectory, its velocity vector can be assumed to have only a tangential component. The tangential component of a particle on a circle is equal to $\xi_i\dot{\varphi}_i$, where $\xi_i$ is the radius of rotation of that particle. Thus, $\left\Vert v_i \right\Vert = \xi_i\dot{\varphi}_i$. 

Throughout the paper, we assume that the particles have unit speed. Therefore, the radius for the coherent group can be estimated approximately as
\begin{equation}
\label{eq:rot_radius_vs_sigma}
	\xi(\sigma) \,\approx\, (\sigma \gamma \sin\alpha)^{-1} \,\,,
\end{equation}
where $\gamma = \left< \gamma \right>_{N_c,t}$ is the average of $\gamma_i$ with respect to the group of coherent particles $N_c$ and with respect to one rotational cycle of that group. The dependence of $\xi$ on $\sigma$ can be seen from Fig.~\ref{fig:localization_traits}(c) obtained directly from simulations and also obtained by the given approximation. We see an agreement of both these methods, which supports the validity of the approximation. It should be indicated that the averaged parameter $\gamma$ is not a constant in \eqref{eq:rot_radius_vs_sigma} but depends on $\sigma$ itself. It is because the coupling strength $\sigma$ influences the shape and size of the localized group, thus, controlling $N_c^i$ and $\gamma_i$ of each particle.

Particles that are not in the coherent group are uniformly distributed over the whole space (Fig.~\ref{fig:localization_traits}(a), dark uniform background) exhibiting a kind of chaotic itinerancy (Fig.~\ref{fig:localization_traits}(b)). Such particles periodically try to follow the circular rotations of the coherent group but generally fall off after some time and continue wandering around. 
It appears that the synchronized group does not consist of the same set of particles all the time. Chaotic particles influence that group in a destabilizing manner (explanation provided in Appendix B), forcing some particles to leave it. But in addition, the chaotic cluster as a whole tries to synchronize, thus introducing new particles into the coherent group. As a result, over time, there are particles that leave and join that group (Fig.~\ref{fig:localization_traits}(d)). This behavior leads to small fluctuations in the group size. The fine balance between escape and capture phenomena enables existence of chimeras for an indefinite time.

\begin{figure}[!b]
	\centering
	\includegraphics[width=0.5\textwidth]{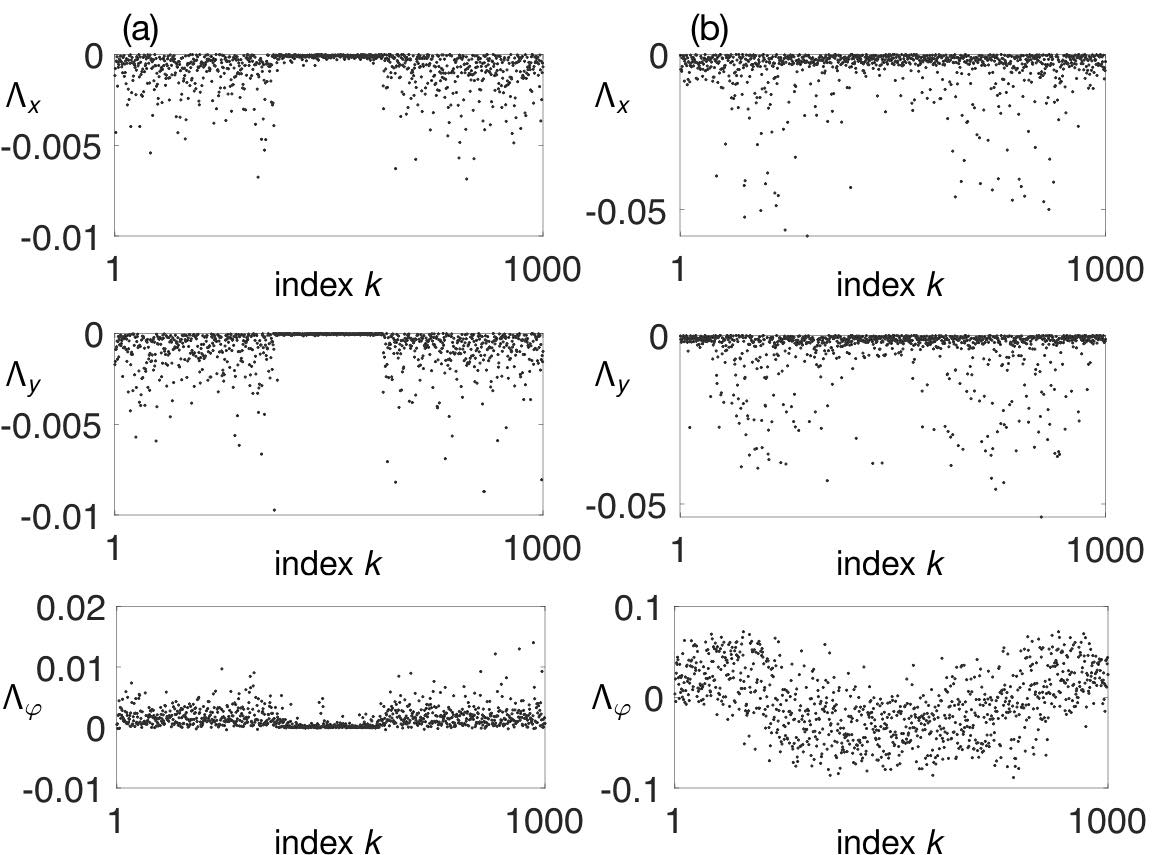}
	\caption
	{
		\label{fig:lyapunov_exponents} Local Lyapunov exponents.
		(a) Local Lyapunov spectra $\Lambda_x$,$\Lambda_y$, and $\Lambda_\varphi$ in the LC case for all particles for $x$,$y$,$\varphi$ variables, respectively. The exponents were computed after $100$ time units after the chimera has been formed. Particles are ordered with respect to $\left<\dot{\varphi}\right>$ in the same manner as in Fig.~\ref{fig:space_time_plots}(a). Other parameters are $\sigma=1.0, \rho=0.3, \alpha=1.54$.
		(b) Local Lyapunov spectra $\Lambda_x$,$\Lambda_y$, and $\Lambda_\varphi$ in the NLC case for all particles for $x$,$y$,$\varphi$ variables, respectively. The exponents were computed after $10$ time units after the chimera has been formed. Particles are ordered with respect to $\left<\dot{\varphi}\right>$ in the same manner as in Fig.~\ref{fig:space_time_plots}(c). Other parameters are $\sigma=1.0, \rho=0.03, \alpha=1.5$.
	}
\end{figure}

We computed the local Lyapunov spectra $\Lambda_x$,$\Lambda_y$, and $\Lambda_\varphi$ for both chimera types (Fig.~\ref{fig:lyapunov_exponents}(a) for LC, Fig.~\ref{fig:lyapunov_exponents}(b) for NLC) to confirm that such dynamics are truly chaotic. The particles' indexes are ordered with respect to $\left<\dot{\varphi}\right>$ in accordance with Fig.~\ref{fig:space_time_plots}(a),(c), respectively. In case of LC, the values of $\Lambda_x$ and $\Lambda_y$ for the synchronized particles remain around 0 and the values of $\Lambda_\varphi$ are slightly negative. For the rest of the particles, first two spectra are negative while the third one is mostly positive. Thus, since the spectrum contains positive exponents, the nature of the system is chaotic. In case of NLC, $\Lambda_x$ and $\Lambda_y$ hardly show any distinction between synchronized and desynchronized groups of particles. The directional spectrum $\Lambda_\varphi$ clearly shows negative values for the synchronized cluster and positive values for the chaotic cluster but it has a high variation. Such a high variation can be explained by the fact that both clusters occupy all the space and influence mutual dynamics to a great extent. By the same reasoning, the system is chaotic in this regime too.

\section{Summary statistics}

One of the important characteristics to describe collective motion in coupled systems is the complex order parameter \cite{pikovsky2003} $Z(t) = R(t) \exp(i\Theta(t)) = 1/N \sum_{j = 1}^{N} \exp(i\varphi_j(t)),$ where $R(t)$ and $\Theta(t)$ can be considered to represent the magnitude and phase of the averaged particle velocity, respectively. The magnitude allows to learn the extent of polarization in the system.

\begin{figure}[!t]
	\centering
	\includegraphics[width=0.5\textwidth]{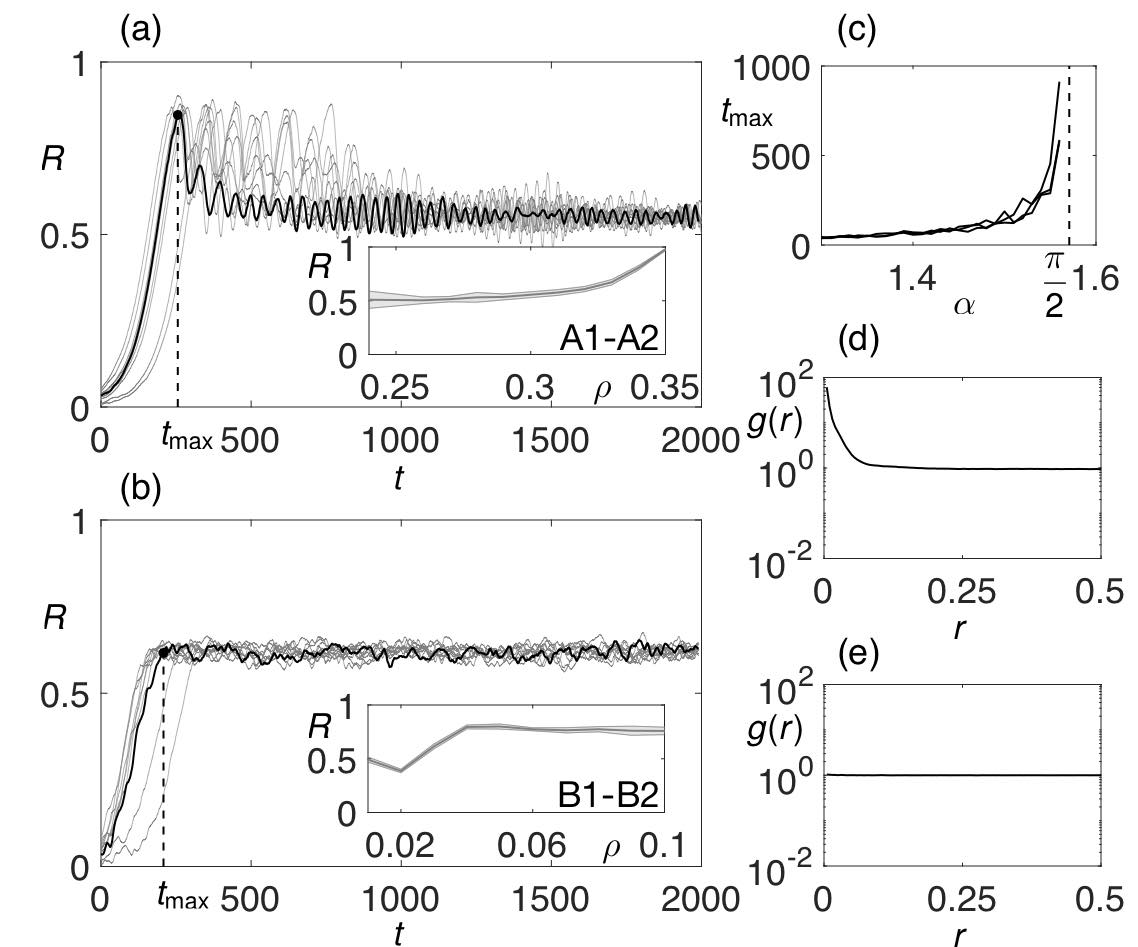}
	\caption
	{
		\label{fig:summary_statistics} Summary statistics for chimera states.
		Evolution of the order parameter magnitude $R$ over time for (a) LC 
		and (b) NLC. Temporal averaging of $t = 10$ is applied. Insets indicate changes in the synchronization level along the lines A1-A2 and B1-B2 from Fig.~\ref{fig:phase_diagram}. 
		(c) The time of maximum synchronization quantified through $R$ versus phase lag $\alpha$, for different radii $\rho\in\{0.1,0.3,0.5\}$.
		(d),(e) Pair distribution functions for LC and NLC, respectively.
		Other parameters are the same as in Fig.~\ref{fig:space_time_plots}.
	}
\end{figure}

Starting from random initial conditions, particles always begin to synchronize. This behavior is well observable from the evolution of $R(t)$ (cf. Fig.\ref{fig:summary_statistics}(a),(b)). The polarization of the system continues till a certain point after which it either decreases or remains at the same level approximately. The time point when it occurs is called the point of maximum synchronization $t_\text{max}$. The stages of the system evolution for both chimera types, described in the previous section, can be observed here additionally, with $t_2=t_\text{max}$. 

The qualitative difference of the order parameter dynamics between the two scenarios is the following. For LC, there is always a pronounced peak at $t_\text{max}$. But shortly afterwards, it drops. This happens when the big disk-shaped group shrinks into a dense spot. It can be seen that it is not the case with NLC where the maximum synchronization is preserved at the same level approximately. It is also worthwhile to notice that when either chimera reaches its stationarity, it is subsequently impossible to differentiate them just by considering the order parameter $Z(t)$ (the insets of Fig.~\ref{fig:summary_statistics}(a),(b) additionally show that the ranges of order parameter magnitudes at stationarity for both chimera types intersect).

\begin{figure}[!b]
	\centering
	\includegraphics[width=0.5\textwidth]{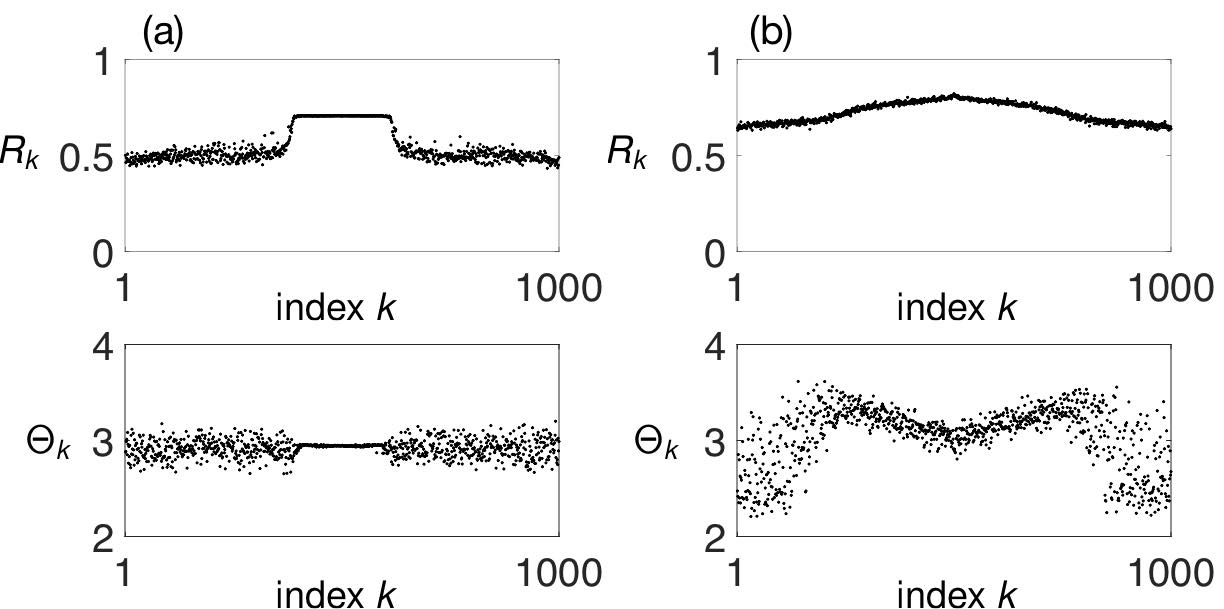}
	\caption
	{
		\label{fig:local_complex_mean_field}
		Local complex mean field $Z_{k}(t) = R_k(t)\exp(i\Theta_k(t))$ at $t=1000$ for (a) LC and (b) NLC. The ordering of particles is the same as in Fig.~\ref{fig:space_time_plots}(a),(c), respectively. Parameters: (a) $\sigma=1.0, \rho=0.3, \alpha=1.54$; (b) $\sigma=1.0, \rho=0.03, \alpha=1.5$.
	}
\end{figure}

A more detailed description to resolve this problem would be to introduce a local complex mean field \cite{xie2014} $Z_{k}(t) = R_k(t)\exp(i\Theta_k(t)) = 1/\left\vert B_{\rho}^{k} \right\vert \sum_{j \in B_{\rho}^{k}} \exp(i\varphi_j(t))$, which is now space-dependent (cf. Fig.~\ref{fig:local_complex_mean_field}). To distinguish the dynamics is now easy. The LC local complex mean field shows explicitly that a synchronized group exists and its presence gives the extreme polarization around itself (the plateau in Fig.~\ref{fig:local_complex_mean_field}(a)). Since NLC does not possess any localization properties by definition, such a plateau is not possible by considering $Z_k(t)$ in this case. 
The drawback of the local complex mean field as a summary statistic is that it contains information about all particle positions and, thus, does not effectively reduce the state space of the problem.

Now, we want to delve deeper into what role each parameter in the model \eqref{eq:spc_sys} plays. First, as it has been seen, the coupling strength $\sigma$ regulates the speed of particle rotations and consequently the radius of such rotations (cf. equation \eqref{eq:rot_radius_vs_sigma}). Its increase facilitates faster system polarization and, thus, smaller $t_\text{max}$. Second, an increase in the radius of interaction $\rho$ leads to an amplification of polarization as more and more particles are engaged into a synchronous group. It is noticeable that for different chimeras the functional dependence $R(\rho)$ differs (see insets in Fig.~\ref{fig:summary_statistics}(a),(b)). We did not observe a considerable influence of $\rho$ upon $t_\text{max}$. Third, an increase of the phase lag $\alpha$ prolongs $t_\text{max}$ in an exponential way independently of $\rho$ (Fig.~\ref{fig:summary_statistics}(c)), with $t_\text{max}\rightarrow\infty$ as $\alpha\rightarrow\pi/2$. Not surprisingly, its increase also decreases system polarization. This looks natural if we look at both chimeras as transient phenomena between complete synchronization (Fig.~\ref{fig:phase_diagram}, region on the left) and chaos (region on the right).

From the definition of both chimera types, the basic difference between LC and NLC is the degree of homogeneity of the system on a small scale. An appropriate function that captures the spatial structure of a system is the pair distribution function $g(r)$. It is a measure of local spatial ordering. It is defined as 
$
g(r) = 1/(\pi r^2 \rho)\left< \sum_{j(\neq i)}\delta(r-r_{ij}) \right>_i
$
, where $r$ is the distance at which the density is to be computed, $\rho=N/L^2$ is the average number density of $N$ particles in the system, $\left< \dots \right>_i$ denotes taking an average over all particles. In Fig.~\ref{fig:summary_statistics}(d),(e) we show the shape of $g(r)$ for both chimeras. For the localized one, it has a very high peak at small $r$ which reflects the fact that there is a localized dense group of particles. For the non-localized one, the peak is absent implying the lack of any localization.

\section{Generalizations}

The presented chimera states obtained from integrating equation \eqref{eq:spc_sys} are not restricted only to that model. To show this, we have also investigated extended versions of the model.

The first important generalization of the chimera model in the context of SPP systems is to introduce noise \cite{yates2009}. Let the particles obey the Langevin equation
\begin{equation}
\label{eq:spc_sys_with_noise}
\mathrm{d}\varphi_{i} \,=\, \frac{\sigma}{\left\vert B_{\rho}^{i} \right\vert} \sum_{j \in B_{\rho}^{i}}  \sin(\varphi_{j} - \varphi_{i} - \alpha)\mathrm{d}t + \sqrt{2D_\varphi}\mathrm{d}W_i,
\end{equation}
where the last additional term represents the noise; $W_i$ and $W_j$ are independent Wiener processes for $i\neq j,\,\, i,j\in\{1,...,N\}$, and $D_\varphi$ is the noise intensity.

\begin{figure}[!t]
	\centering
	\includegraphics[width=0.5\textwidth]{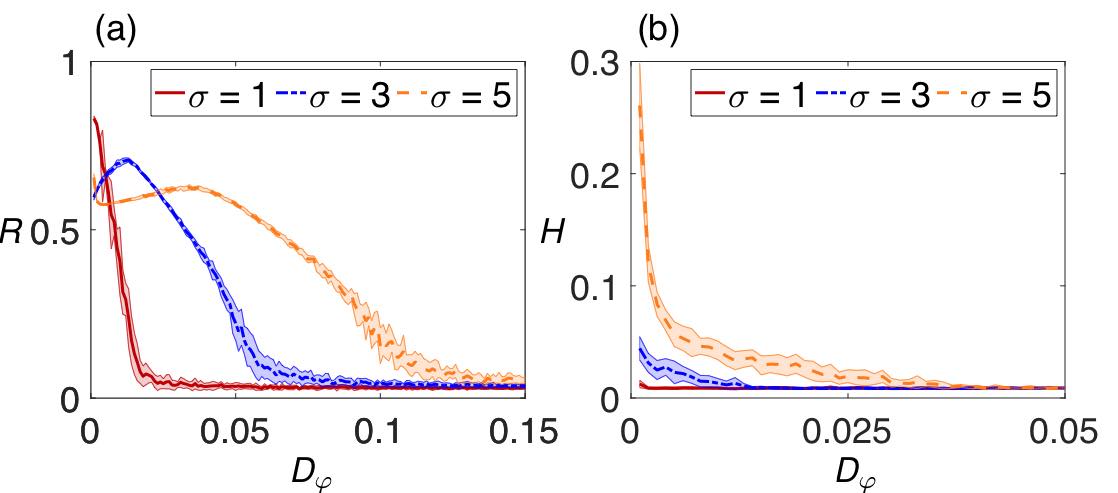}
	\caption
	{
		\label{fig:summary_statistics_stochastic} Summary statistics for chimera states under noise.
		(a) Values of global order parameter magnitude $R$ versus noise intensity $D_\varphi$ for solutions of equation \eqref{eq:spc_sys_with_noise}. Different colors (line styles) correspond to different coupling coefficients $\sigma$; (b) Localization measure $H$ versus noise intensity $D_\varphi$. A shaded region shows the standard deviation along a curve. Other parameters are $\rho=0.3, \alpha=1.54, N=1000$.
	}
\end{figure}

The summary of the system's dynamics is presented in Fig.~\ref{fig:summary_statistics_stochastic} through the magnitude $R$ of the order parameter as a measure of polarization and through $H=\int \left| g(r)-1 \right| \mathrm{d}r$ as a global measure of localization (see another usage of it in Appendix A). The addition of noise prevents particles from gathering into dense formations for small $\sigma$ (cf. Fig.~\ref{fig:summary_statistics_stochastic}, $\sigma=1$) but does not prevent the partial synchronization (i.e., LC turns into NLC). As $\sigma$ is increased, we recover again the LC states ($\sigma=5$, intermediate $R$ and high $H$). There is another interesting phenomenon. If $D_\varphi$ increases, the order parameter magnitude $R$ initially grows. This occurs because the higher the value of $D_\varphi$ is, the more spacious the localized group becomes. The process continues till the point at which the order parameter is maximized and the localization measure reaches a minimal possible value. At this point, the system can again be described as NLC. Further increase of $D_\varphi$ merely destroys the remaining phase synchronization.

In order to confirm that the reported self-propelled chimera states are not the result of applying only the piece-wise constant interaction kernel in equation \eqref{eq:spc_sys}, we have considered two other types of couplings commonly applied for the Kuramoto model, i.e. the cosine and exponential couplings. However, in order to embed such a coupling function into our system, we need to rewrite it the following way:
\begin{align*}
\dot{r}_{i} \,&=\, v(\varphi_{i}),\qquad \\
\dot{\varphi}_i \,&=\, \frac{\sigma}{\left\vert B_{\rho}^{i} \right\vert} \sum_{j \in B_{\rho}^{i}}  \sin(\varphi_{j} - 
\varphi_{i} - \alpha) \\
&= \sigma \frac{\sum\limits_{j = 1}^N H(\rho - \left\Vert r_j - r_i \right\Vert) \sin(\varphi_j - \varphi_i - \alpha)}{\sum\limits_{j = 1}^N H(\rho - \left\Vert r_j - r_i \right\Vert)},
\end{align*}
where $H$ is a Heaviside step function such that $H(x) = \begin{cases} 0, & x<0,\\ 1, & x\geq0. \end{cases}$ This model is an equivalent representation of the model \eqref{eq:spc_sys}. Now we replace the piecewise-constant Heaviside step function with a general kernel function which leads to the following model
\begin{equation}
\label{eq:spc_sys_with_kernel}
\begin{aligned}
\dot{r}_{i} \,&=\, v(\varphi_{i}),\qquad\\
\dot{\varphi}_{i} \,&=\, \sigma \frac{\sum_{j = 1}^N G(\left\Vert r_j - r_i \right\Vert) \sin(\varphi_j - \varphi_i - \alpha)}{\sum_{j=1}^N G(\left\Vert r_j - r_i \right\Vert)},
\end{aligned}
\end{equation}
where $G$ is a distance-dependent kernel function that provides non-local coupling between the particles.

One common choice for the kernel function in the systems of coupled oscillators is
\begin{equation}
\label{eq:cos_extension}
G(r) = 1 + A \cos(2 \pi r),
\end{equation}
where $0 \leq A \leq 1$ \cite{abrams2004} is a tunable parameter.

\begin{figure}[!b]
	\centering
	\includegraphics[width=0.5\textwidth]{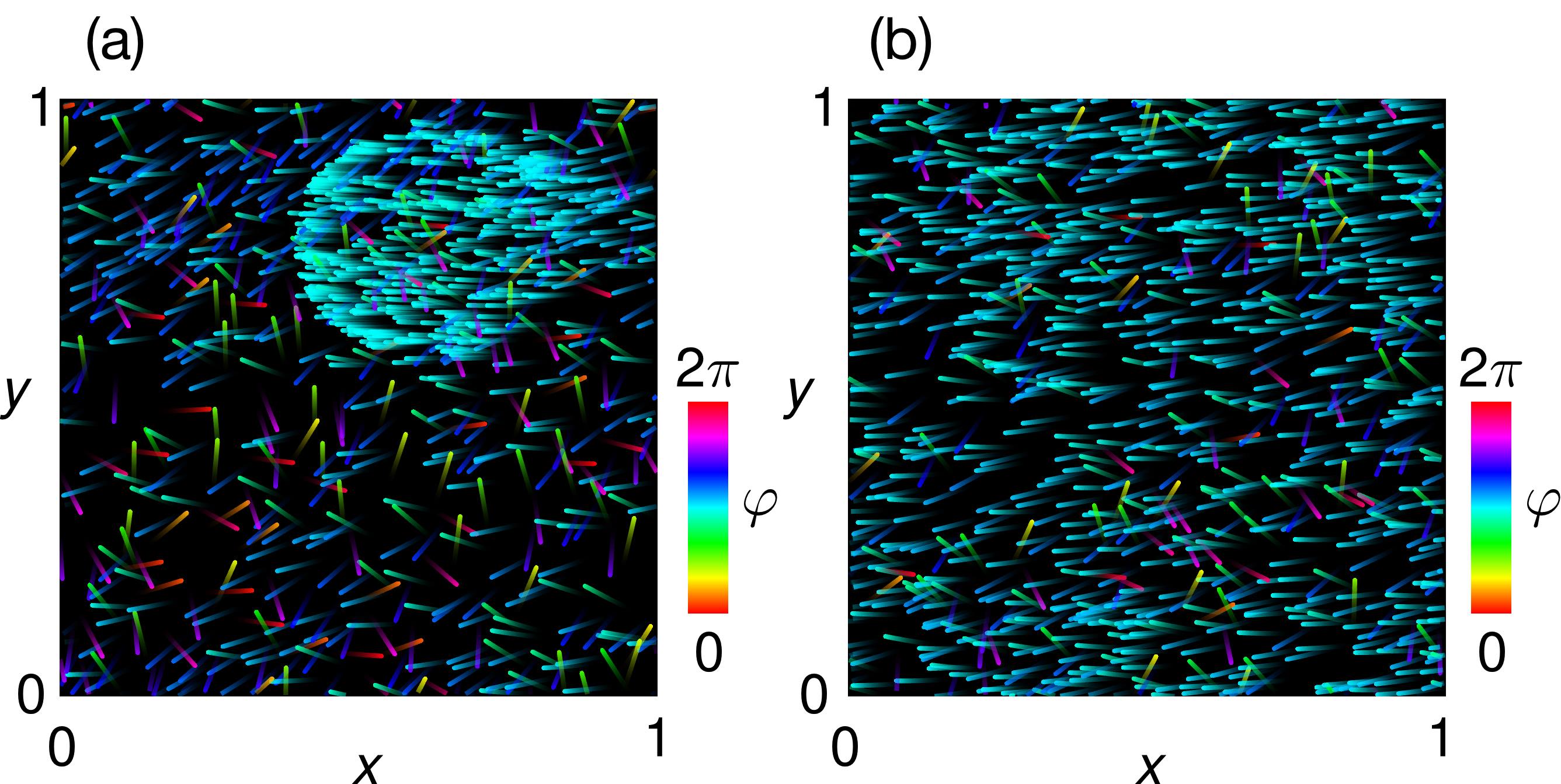}
	\caption
	{
		\label{fig:cos_kernel}
		Chimera states due to the generalized model \eqref{eq:spc_sys_with_kernel} with the cosine kernel \eqref{eq:cos_extension}. (a) Localized chimera state with parameters $\sigma=1.0,\alpha=1.53, A=1.0$; (b) Non-localized chimera state with parameters $\sigma=1.0, \alpha=1.56, A=1.0$.
	}
\end{figure}

It appears that the modified system \eqref{eq:spc_sys_with_kernel} also produces various chimeric patterns (cf. Fig.~\ref{fig:cos_kernel}). We have found that in the LC case, synchronized particles form a localized structure but it has a form of a ball rather than a spot. The dynamics in the NLC case is qualitatively similar to the corresponding behavior due to equation \eqref{eq:spc_sys}. 

\begin{figure}[!t]
	\centering
	\includegraphics[width=0.5\textwidth]{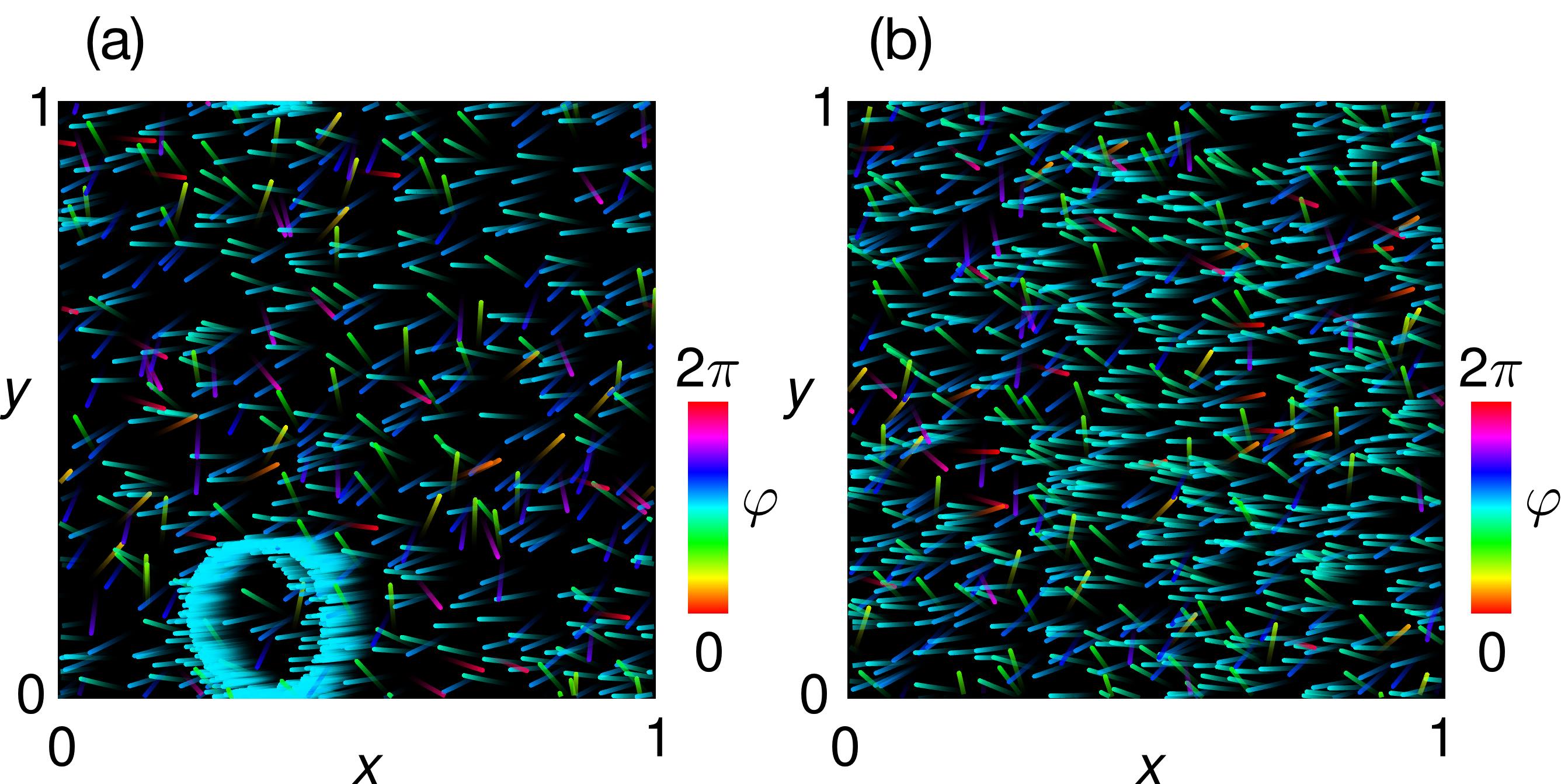}
	\caption
	{
		\label{fig:exp_kernel}
		Chimera states due to the generalized model \eqref{eq:spc_sys_with_kernel} with the exponential kernel \eqref{eq:exp_extension}. (a) Localized chimera state with parameters $\sigma=1.0, \alpha=1.54, k=4.0$; (b) Non-localized chimera state with parameters $\sigma=1.0, \alpha=1.54, k=20.0$.
	}
\end{figure}

Another common choice for the kernel function is
\begin{equation}
\label{eq:exp_extension}
G(r) = \frac{k}{2} e^{- k r},
\end{equation}
where $k$ is a tunable parameter \cite{kuramoto2002}.
Considering the inclusion of the exponential coupling into the extended system \eqref{eq:spc_sys_with_kernel}, two different states of chimeras can again be observed (cf. Fig.~\ref{fig:exp_kernel}). Remarkably, due to the global connection of the particles, the LC state gains additional peculiarity. As soon as the localized group has appeared, it also synchronizes particles outside of the group, thus creating a subsequently synchronized scattered group. The resulting coherent cluster consists of both localized and non-localized particles. The change of $\alpha$ leads to the alteration of the shape of the localized group, e.g., the higher it is the less concentrated and convex-shaped the group becomes. With increase of the tuning parameter $k$, the size of the localized cluster decreases and as a result, the number of scattered but synchronized particles grows. Again, the dynamics in the NLC case is qualitatively similar to the corresponding behavior due to equation \eqref{eq:spc_sys}.

\section{Continuum limit}
In this section we derive the continuum limit for the Langevin equation \eqref{eq:spc_sys_with_noise} and show by numerical integration of the resulting 3+1-dimensional partial differential equation that chimeras are also preserved in this limit.

We follow the approach of \cite{dean:phys_a}, which was also used in \cite{grossmann2014} for the derivation of the Fokker-Planck equation from a system of self-propelled particles. We define the microscopic density function for $N$ particles
$
f^N(r,\varphi,t)=1/N\sum^{N}_{i=1}\delta(r_i(t)-r)\delta(\varphi_i(t)-\varphi),
$
which includes all the particle state variables. Using Ito's calculus (see the detailed derivation in Appendix C), we obtain the following closed-form equation for the time evolution of the density function
\begin{equation} 
\label{eq:emperical probability density function}
\begin{aligned}
\partial_t &f^N(r,\varphi,t) = -\nabla_r\cdot\left[ f^N(r,\varphi,t) \dot{r}(\varphi,t) \right] - \\
&\partial_{\varphi}\left[ f^N(r,\varphi,t) \dot{\varphi}(r,\varphi,t) \right] + D_\varphi\partial_{\varphi\varphi}f^N(r,\varphi,t),
\end{aligned}
\end{equation}
where $f^N\dot{r}$ is the flux due to the motion of particles and $f^N\dot{\varphi}$ is the angular flux resulting from the alignment mechanism. Since $f^N$ contains the information of all the particles' state variables,
equation \eqref{eq:emperical probability density function} is of the Klimontovich type \cite{nicholson:wiley}.

Under the molecular chaos assumption \cite{spohn:springer}, which corresponds to neglecting all pre-collisional particle correlations, we arrive at the mean-field limit \cite{chepizhko:2014} as the number of particles $N$ goes to infinity. 
In this limit, the microscopic density function converges to a one-particle density function  $f=f(r,\varphi,t) = \lim_{N\rightarrow \infty}  f^N({r},\varphi,t)$ \cite{spohn:springer, carrillo:2014cdbc}, which is finally independent of the explicit particle information. If we express variables $x,y,\varphi$ explicitly, we finally arrive at the following Fokker-Planck equation with a non-local coupling term
\begin{equation}
\label{eq:one-particle probability density function}
\begin{aligned}
&\partial_t f = D_\varphi \partial_{\varphi\varphi}f -\partial_x(f \cos\varphi) -\partial_y(f \sin\varphi) - \\
& \partial_\varphi \left[f \frac{\sigma}{\left\vert B_\rho \right\vert} \iint\limits_{B_\rho}\int\limits_{0}^{2\pi} f' \sin\left( \varphi' - \varphi - \alpha \right) \mathrm{d}x'\mathrm{d}y'\mathrm{d}\varphi' \right],
\end{aligned}
\end{equation}
where $\left\vert B_\rho \right\vert$ represents the normalization term in the form of the neighborhood mass 
\begin{equation*}
\lvert B_\rho \rvert = \iint\limits_{B_\rho}\int\limits_{0}^{2\pi} f \mathrm{d}x'\mathrm{d}y'\mathrm{d}\varphi'
\end{equation*}
and $f'=f(x',y',\varphi')$.

\begin{figure*}[!t]
	\centering
	\centerline{\includegraphics[width=1\textwidth]{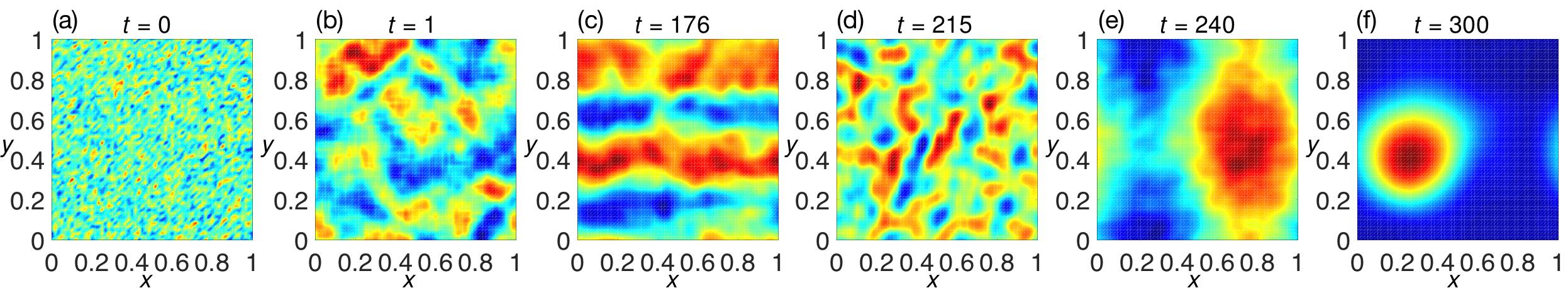}}
	\caption
	{
		\label{fig:projected_pde}
		Solution for the 3+1-dimensional density function from the continuum limit \eqref{eq:one-particle probability density function}, projected into spatial coordinates $(x,y)$. The projection was done so as to ensure $f(x,y,t)=\int_{0}^{2\pi} f(x,y,\varphi,t) \mathrm{d}\varphi$. The color corresponds to the density value, normalized per each frame, so that all dynamical variations are captured. The complete dynamics of $f(x,y,\varphi,t)$ and the original density scale are provided in Fig.~\ref{fig:complete_pde}.
		The solution here is shown as a sequence of characteristic phases that the system passes (see their description in the text).
	}
\end{figure*}

Despite the fact that the dynamics of \eqref{eq:one-particle probability density function} depends on $x$, $y$, and $\varphi$, here we present only the projections of the solution into $(x,y)$ because it allows simpler representation while still keeping the key aspects of the dynamics (the corresponding complete dynamics is described in Appendix C; see videos S4 and S5 under \cite{supp_mat,bcs_youtube_channel} for the temporal dynamics).
Although our emphasis is primarily on the most fascinating regime that is equivalent to the LC state of the model \eqref{eq:spc_sys}, the other regimes also exist in the continuum limit. 

At the beginning of a simulation, each grid point is initialized uniformly with small perturbation and then rescaled so as to keep the overall system density normalized (cf. Fig.~\ref{fig:projected_pde}(a)). As the system starts to move, many patches of high density appear and propagate according to the given velocity field (cf. Fig.~\ref{fig:projected_pde}(b)). Those patches first merge into tubes of high density, which, in turn, agglomerate into a thick layer, uniform in $(x,y)$. 
The layer moves along the $\varphi$ direction for a considerable amount of time, slowly shrinking. At some point, the layer shrinks rapidly and wavy structures appear in front of it (cf. Fig.~\ref{fig:projected_pde}(c)). Soon, those structures become irregular (cf. Fig.~\ref{fig:projected_pde}(d)) and the layer transforms into an elongated object of high density (cf. Fig.~\ref{fig:projected_pde}(e)). Eventually the object condenses into a small ellipsoidal shape that moves along a helical trajectory (cf. Fig.~\ref{fig:projected_pde}(f)). Also note that the final localized high-density shape coexists with the surrounding of non-zero mass. 
Because of that, the system state resembles the motion of a coherent group of particles through an incoherent surrounding, i.e., the LC. 

\begin{figure}[!b]
	\centering
	\includegraphics[width=0.5\textwidth]{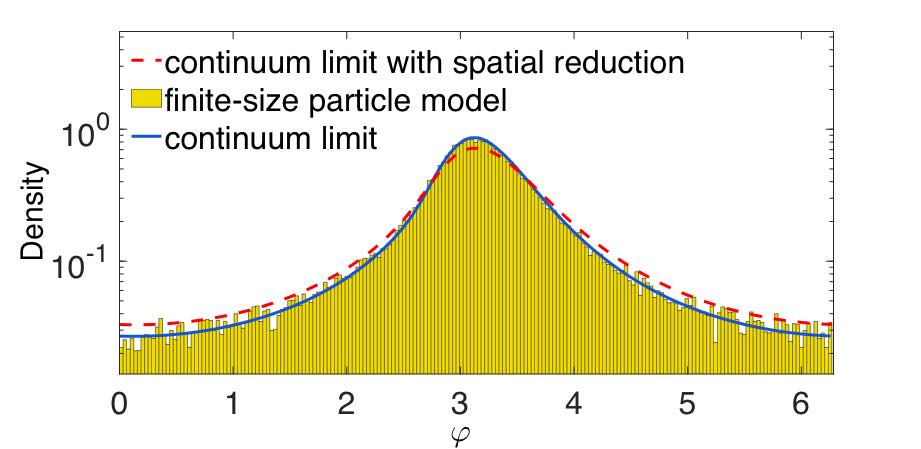}
	\caption{Comparison of the dynamics in the case of the non-localized chimera state from different models: (i) yellow histogram of partiles' directions is given by the solution of a finite-size particle model \eqref{eq:spc_sys_with_noise} with the number of particles $N=5\cdot 10^4$; (ii) solid blue curve represents the projection of the solution for a 3+1-dimensional partial differential equation \eqref{eq:one-particle probability density function} into $\varphi$ coordinates; (iii) dashed red curve represents the solution of a reduced partial differential equation \eqref{eq:angle evolution}. Parameters for the three models are $\sigma=1.0, \rho=0.03, \alpha=1.5, D_\varphi=0.01$.}
	\label{fig:pde_comparison}
\end{figure}

If we consider the NLC behavior, the final solution consists of a thin layer, which moves along the $\varphi$ direction and which is uniform in $x,y$. In this case, we can employ the spatial homogeneity that is expressed as $f(x,y,\varphi,t) = \hat{f}(\varphi,t)$ for the considered unit square domain. Since $\hat{f}$ does not depend on $x,y$ anymore, we can drop the first two terms on the right hand side of \eqref{eq:one-particle probability density function}. Integration of the remaining equation over the unit square domain yields
\begin{equation}
\label{eq:angle evolution}
\begin{aligned}
&\partial_t\hat{f}(\varphi,t) = D_\varphi\partial_{\varphi\varphi}\hat{f}(\varphi,t) \\
&- \partial_\varphi \left[ \hat{f}(\varphi,t)\frac{\sigma}{\left|B_\rho\right|} \int\limits_{0}^{2\pi}\hat{f}(\varphi',t)\sin(\varphi'-\varphi-\alpha) \mathrm{d}\varphi' \right],
\end{aligned}
\end{equation}
where the normalization term reads
\begin{equation*}
\lvert B_\rho \rvert = \int\limits_{0}^{2\pi} \hat{f}(\varphi',t) \mathrm{d}\varphi'.
\end{equation*}
The solution to this equation exhibits a pronounced peak for a certain phase and significant non-zero mass at other phases (cf. Fig.~\ref{fig:pde_comparison}, the dashed red curve). Such a form corresponds to the NLC solution of the finite-size particle model \eqref{eq:spc_sys_with_noise} in such a way that there is a cluster of synchronized particles but at the same time there is still a cluster of chaotically moving particles. Moreover, as it has been assumed, both clusters are uniformly distributed in space.

Even though the reduced equation does recover the partial synchronization property needed for the system to be described as a chimera, the form of the solution to this reduced equation is not entirely similar than that one from the original 3+1-dimensional continuum limit \eqref{eq:one-particle probability density function}, projected into the $\varphi$-coordinates (cf. Fig.~\ref{fig:pde_comparison}, the solid blue curve). The figure shows that the solution to \eqref{eq:one-particle probability density function} fits well the phase histogram, which is obtained from particles' directions due to the Langevin equation \eqref{eq:spc_sys_with_noise}. At the same time, the solution to \eqref{eq:angle evolution} smooths out the pronounced peak to a higher extent than the original continuum limit density function. It can be explained by the fact that in this case we ignore all the spatial inhomogeneity during the formation of the chimera.

\section{Conclusions}

Our model for chimera states in SPP systems has been built from very simple rules and can easily be generalized to account for various types of interactions. For example, the inclusion of short-range repulsion is essential in order to express the physical interaction in natural environments.  
Previous works about the collective behavior were primarily concerned about the synchronizing nature of interactions. Our research uncovers the important interplay of synchronization and chaos in order to explain previously not discussed self-organized structures, which we termed self-propelled chimeras. We presented two distinct types of chimera states, namely, localized and non-localized. Notably, even though the main emphasis of the paper is on the deterministic nature of the reported states, we have also shown that the addition of noise does not destroy these particular states. Furthermore, we derived the continuum limit for the finite-size particle model and confirmed that chimeras do exist in the limit. 
The found spatio-temporal dynamics are fundamentally different to the previously reported spatio-temporal structures in SPP systems that were caused by stochastic terms. To prove the existence of genuine chimera states in real collectives such as animal groups will be challenging and will require dedicated experiments where the strength of fluctuations can be well controlled.

\section*{Acknowledgement}

We are very grateful to Lutz Schimansky-Geier and Mark Timme for helpful discussions. Y.M. acknowledges support and hospitality of TU Darmstadt within the KIVA framework. We also gratefully acknowledge the computing time provided by the Hessian Competence Center of High Performance Computing on the Lichtenberg High Performance Computer at TU Darmstadt.

\section*{Appendix A: Numerical methods}

All the simulations of equation \eqref{eq:spc_sys} were performed with the Runge-Kutta-4 integration scheme with $\Delta t\in[0.001;0.01]$. The simulations of the Langevin equation \eqref{eq:spc_sys_with_noise} were performed with the Stochastic Runge-Kutta integration method (strong order 1.5 Taylor scheme) \cite{platen}. The simulations of the continuum limit partial differential equation \eqref{eq:one-particle probability density function},\eqref{eq:appendix:one-particle probability density function_transformed} were performed with the Sweby's flux-limited method \cite{laney:comp_gas_dyn}, which uses numerical wave speed splitting. Additionally, that method was used together with the enforcement of the nonlinear stability condition called the positivity condition \cite{laney:comp_gas_dyn}. We used the Sweby's method with the superbee flux limiter, the time-step was varied as $\Delta t\in[10^{-4};10^{-3}]$, the spatial discretization for each dimension was varied as $\Delta x,\Delta y, \Delta \varphi \in [\frac{1}{60};\frac{1}{20}]$. The reduced partial differential equation \eqref{eq:angle evolution} was integrated using the MacCormack's technique \cite{anderson:comp_fluid_dyn} together with the addition of dissipative terms \cite{davis_siam}; the domain discretization was varied as $\Delta \varphi \in [\frac{2\pi}{512};\frac{2\pi}{128}]$. We would also like to indicate here that there are other frameworks to numerically solve Boltzmann-type equations for self-propelled particle systems \cite{thuroff:prx,ihle:pre}. The Lyapunov spectra were calculated using the Gram-Schmidt reorthonormalization procedure (with the period of reorthonormalization $t = 1$ time unit). Most of the programs were written in C++, some of them with the usage of MPI or CUDA for parallelization and with the usage of OpenGL for graphics rendering; some analysis and the majority of the plots were rendered in MATLAB.

\subsection*{Phase diagram construction}

The phase diagram in Fig.~\ref{fig:phase_diagram} was produced with the help of a continuation method. First, we used $\alpha=1.54, \rho=0.3$ as a starting point for the diagram generation, and its final state, when a localized chimera had been created, as an initial condition for the next parameter tuples. We chose them to be $\alpha\pm\Delta\alpha, \rho\pm\Delta\rho$, with $\Delta\alpha=0.005, \Delta\rho=0.01$. We initially integrated equation \eqref{eq:spc_sys} for each new parameter tuple over $t=1000$ time units, and then continued the integration over $t=100$ time units gathering summary statistics, i.e. the global order parameter and the pair distribution function. For a tuple to belong to the LC region, we imposed the following conditions:
\begin{enumerate}
	\item the mean value of the order parameter magnitude did not synchronize or desynchronize completely $\left< R(t) \right>_t \in (0;1)$;
	\item the standard deviation of the order parameter magnitude was $<0.25$ so that the solution did not deviate much;
	\item the pair distribution function $g(r)$ (see its definition in the main text) for LC shows a considerable peak for small $r$. This behavior is expected since it reflects the fact that there is a highly dense group of particles. Note that $g(r)$ has non-zero values starting directly from $r=0$ because equation \eqref{eq:spc_sys} does not dictate any short-range repulsion between particles. As $r$ increases, the homogeneity scale overcomes quickly ($g(r)=1$) since the group size is limited. For NLC, $g(r)$ does not have any peaks which is expected since any localization here is supposed to be absent. As a result, the pair distribution function is always on the homogeneity scale for NLC. Consequently, we calculate an $L^1$-norm $H=\int \left| g(r)-1 \right| \mathrm{d}r$ as a condition for a system to possess the localization property. Namely, it should be $>0.1$ in order to be characterized as LC.
\end{enumerate}
If all three conditions were satisfied, the tuple $(\alpha,\rho)$ was accepted as the LC-tuple and the next tuples were generated. The program execution continued until there were no more tuples to analyze in the domain of interest. An initial condition for each new tuple was taken from the final state of the system integrated with the closest LC-tuple. And the process continued in the same manner as it had been for the initial tuple.

For NLC, the initial tuple was $\alpha=1.5, \rho=0.03$, the standard deviation of the order parameter was bounded with $0.1$, the $L^1$-norm was imposed to be $<0.02$.

\section*{Appendix B: Interdependence of localized and non-localized particle subpopulations in a localized chimera state}

The LC state comprises two clusters: one with synchronized particles and the other one with chaotic particles. It appears that the existence of one of the clusters is necessary for the other one to function. We can show this by extracting one cluster out of the system and analyzing how the remaining particles behave.

First, if we remove the cluster of synchronized particles, the system will be left with only chaotic particles. Thereby, we reduce the problem to the description of the system initialized with random initial conditions. As it is discussed throughout the paper, if the number of particles is still higher than some critical value, the system again evolves into the LC state.

Second, if we extract the cluster of chaotic particles, the system becomes completely polarized. We can see this from equation \eqref{eq:spc_sys}:
\begin{equation}
\label{eq:in_phase}
\dot{r}^* = \begin{pmatrix} \cos\varphi^* \\ \sin\varphi^* \end{pmatrix},\qquad \dot{\varphi}^* = -\sigma\sin\alpha,
\end{equation}
where $\varphi^*=\varphi^*(t)$ for each particle (but $r^* = r^*(t)$ is different for each one). Such a synchronized state is linearly unstable and only occurs when we choose a special initial condition, where every particle is synchronized. 
The fact that the in-phase solution, which satisfies \eqref{eq:in_phase}, is unstable, we can show via the linearized Poincar\'e map \cite{strogatz}. The solution is given by: $\varphi_1 = \varphi_2 = \dots = \varphi_N = \varphi^*(t)$. To determine the stability, we put $\varphi_i(t) = \varphi^*(t) + \eta_i(t)$, where $\eta_i(t)$ are infinitesimal perturbations. Let's substitute the perturbed solutions $\varphi_i$ into equation \eqref{eq:spc_sys} and linearize it in $\eta$. We obtain
\begin{align*}
\dot{x}_i &\approx \cos\varphi^* - \eta_i\sin\varphi^*, \\
\dot{y}_i &\approx \sin\varphi^* + \eta_i\cos\varphi^*, \\
\dot{\eta}_i &\approx -\sigma\eta_i\cos\alpha + \frac{\sigma\cos\alpha}{\left\vert B_\rho^i \right\vert} \sum_{j \in B_\rho^i} \eta_i.
\end{align*}
Let's introduce a new variable $\xi_i = \eta_{i + 1} - \eta_i$. Then its derivative is
\begin{equation*}
\dot{\xi}_i = \dot{\eta}_{i+1} - \dot{\eta_i} = -\sigma(\eta_{i+1}-\eta_i)\cos\alpha = -\sigma\xi_i\cos\alpha.
\end{equation*}
Note that we have used the fact that since all the particles are closely localized, we can consider the equality of neighborhoods for each of them. This system is easily solved and in terms of perturbations we have:
\begin{equation}
\label{eq:dependence_proof_1}
\eta_{i+1}(T) - \eta_i(T) = (\eta_{i+1}(0) - \eta_i(0))e^{2\pi\cot\alpha}.
\end{equation}
Our goal is to prove that the periodic orbit given by $(r^*,\varphi^*)$ is unstable. We will do this by contradiction. Let's assume that all the characteristic multipliers of the periodic orbit are $\forall i \quad \lambda_i < 1$. Then after one cycle, each perturbation is
\begin{equation*}
\forall i \quad \eta_i(T) = \lambda_i \eta_i(0).
\end{equation*}
For each pair of perturbations we can write
\begin{equation}
\label{eq:dependence_proof_2}
\forall i \quad \eta_{i+1}(T) - \eta_{i}(T) = \lambda_{i+1} \eta_{i+1}(0) - \lambda_{i} \eta_{i}(0).
\end{equation}
Let's subtract \eqref{eq:dependence_proof_1} from \eqref{eq:dependence_proof_2}:
\begin{equation*}
(\lambda_i - e^{2\pi\cot\alpha}) \eta_i(0) = (\lambda_{i+1} - e^{2\pi\cot\alpha}) \eta_{i+1}(0).
\end{equation*}
Since we have assumed all $\lambda_i < 1$, and we are interested in the regime with $|\alpha|<\frac{\pi}{2}$, both parentheses are of the same sign. Thus, we have
\begin{equation*}
\forall i \quad \eta_i(0) \eta_{i+1}(0) > 0.
\end{equation*}
But since the infinitesimal perturbations are allowed to be arbitrary, there exists a pair of perturbations of different signs:
\begin{equation*}
\exists i: \eta_i(0) \eta_{i+1}(0) < 0.
\end{equation*}
As a result, we have a contradiction and, thus, we deduce that $\exists i: \left\vert \lambda_i \right\vert > 1$. This in turn implies that perturbations along this direction grow, and so the whole in-phase solution $(r^*,\varphi^*)$ is unstable.

We can now see why there is an exchange of particles between localized and non-localized clusters (see its description in the main text). On one hand, because the localized cluster is on its own an unstable solution, chaotic particles serve as destabilizing perturbations to it, thus, forcing it to lose some of its members. On the other hand, because chaotic particles on their own try to align, a part of them synchronizes with the polarized group, thus, increasing this group's size.

\section*{Appendix C: One-particle density function}

\subsection*{Derivation}

During the derivation, we follow the approach of \cite{dean:phys_a}. For convenience, let's merge all particle state variables into one
\begin{equation*}
\vec{s}_i = (x_i, y_i, \varphi_i).
\end{equation*}
Then the Langevin equation for $\vec{s}$ takes the form
\begin{equation*}
\mathrm{d}\vec{s}_i = \vec{u}_i\mathrm{d}t + \sqrt{2D_\varphi}\mathrm{d}\vec{B}_t,
\end{equation*}
where $\vec{u}_i = \vec{u}(\vec{s}_i) = \begin{pmatrix} \cos\varphi_i \\ \sin\varphi_i \\ \frac{\sigma}{\left\vert B_{\rho}^i \right\vert}\sum_{j\in B_{\rho}^i}\sin(\varphi_j - \varphi_i - \alpha) \end{pmatrix}, \vec{B}_t = \begin{pmatrix} B_t^x \\ B_t^y \\ B_t^{\varphi} \end{pmatrix}$ - three-dimensional Brownian motion. However, since we consider that only the direction of motion $\varphi$ is subject to random environmental effects, we put $B_t^x = B_t^y \equiv 0$.

Initially, let's define the microscopic phase space density function in the following way:
\begin{equation}
\label{eq:global_density_definition}
f^N(\vec{s},t) = \frac{1}{N}\sum_{i=1}^N f_i(\vec{s},t) = \frac{1}{N}\sum_{i=1}^N\delta_i^3(\vec{s},t),
\end{equation}
where $f_i$ represents the density function of a single particle, $\delta_i^3(\vec{s},t)=\delta(x-x_i)\delta(y-y_i)\delta(\varphi-\varphi_i)$, $\delta$ is the Dirac delta function. Consider an arbitrary function from the Schwarz space $g \in \mathcal{S}$. By the main property of the Dirac delta function
\begin{equation}
\label{eq:dirak_property}
\int g(\vec{s})\delta_i^3(\vec{s},t)\mathrm{d}\vec{s} = g(\vec{s}_i(t)),
\end{equation}
where the integration is performed over all of the state variables $\mathrm{d}\vec{s} = \mathrm{d}x\mathrm{d}y\mathrm{d}\varphi$. According to the general Ito formula \cite{oksendal}, the function $g(\vec{s})$ is expressed as
\begin{align*}
\frac{\mathrm{d}g(\vec{s}_i)}{\mathrm{d}t} &= \frac{\partial g(\vec{s}_i)}{\partial t} + [\nabla g(\vec{s}_i)]\cdot\vec{u}_i \\
&+ \sqrt{2D_\varphi}[\nabla g(\vec{s}_i)]\cdot\vec{\eta}_i + D_\varphi\Delta g(\vec{s}_i),
\end{align*}
where the gradient $\nabla$ acts on all particle's variables, $\eta_{i}$ is normally distributed noise at each time point.
Since $g$ does not depend on time explicitly, the first term on the right side vanishes and we then use the property \eqref{eq:dirak_property} to separate the dependence of $g$ on a particle's index $i$. Thus, we have
\begin{align*}
\frac{\mathrm{d}g(\vec{s}_i)}{\mathrm{d}t} &= \int\biggl\{ [\nabla g(\vec{s})]\cdot\vec{u} \\
&+ \sqrt{2D_\varphi}[\nabla g(\vec{s})]\cdot\vec{\eta}_i + D_\varphi\Delta g(\vec{s}) \biggr\}\delta_i^3(\vec{s},t) \mathrm{d}\vec{s},
\end{align*}
where $\vec{u} = \begin{pmatrix} \cos\varphi \\ \sin\varphi \\ \frac{\sigma}{\left\vert B_{\rho} \right\vert}\sum_{j\in B_{\rho}}\sin(\varphi_j - \varphi - \alpha) \end{pmatrix}$.

We use integration by parts together with the divergence theorem in order to rearrange the integration:
\begin{equation}
\begin{aligned}
\label{eq:deriv_1}
&\frac{\mathrm{d}g(\vec{s}_i)}{\mathrm{d}t} = \int\biggl[ -\nabla\cdot(\delta_i^3(\vec{s},t)\vec{u}) \\
- \sqrt{2D_\varphi}&\nabla\cdot(\delta_i^3(\vec{s},t)\vec{\eta}_i) + D_\varphi\Delta\delta_i^3(\vec{s},t) \biggr]g(\vec{s}) \mathrm{d}\vec{s}.
\end{aligned}
\end{equation}
From \eqref{eq:dirak_property}, we also deduce
\begin{equation}
\label{eq:deriv_2}
\frac{\mathrm{d}g(\vec{s}_i)}{\mathrm{d}t} = \int g(\vec{s})\frac{\partial\delta_i^3(\vec{s},t)}{\partial t} \mathrm{d}\vec{s}.
\end{equation}
Let's consider the right hand sides of \eqref{eq:deriv_1}-\eqref{eq:deriv_2}. Since both expressions are considered for arbitrary functions $g(\vec{s})$, we conclude the equality of expressions involving delta functions:
\begin{align*}
\frac{\partial\delta_i^3(\vec{s},t)}{\partial t} &= -\nabla\cdot(\delta_i^3(\vec{s},t)\vec{u}_i) \\
&- \sqrt{2D_\varphi}\nabla\cdot(\delta_i^3(\vec{s},t)\vec{\eta}_i) + D_\varphi\Delta\delta_i^3(\vec{s},t).
\end{align*}
Now, in order to obtain the equation for the density function $f^N(\vec{s},t)$, we average the above equation over the $i$'s and use the definition of this function \eqref{eq:global_density_definition}. We obtain afterwards
\begin{equation}
\begin{aligned}
\label{eq:deriv_3}
&\frac{\partial f^N(\vec{s},t)}{\partial t} = -\nabla\cdot(f^N(\vec{s},t)\vec{u}) \\
- \frac{\sqrt{2D_\varphi}}{N}&\sum_{i=1}^{N}\nabla\cdot(\delta_i^3(\vec{s},t)\vec{\eta}_i) + D_\varphi\Delta f^N(\vec{s},t).
\end{aligned}
\end{equation}

\begin{figure*}
	\centering
	\centerline{\includegraphics[width=1\textwidth]{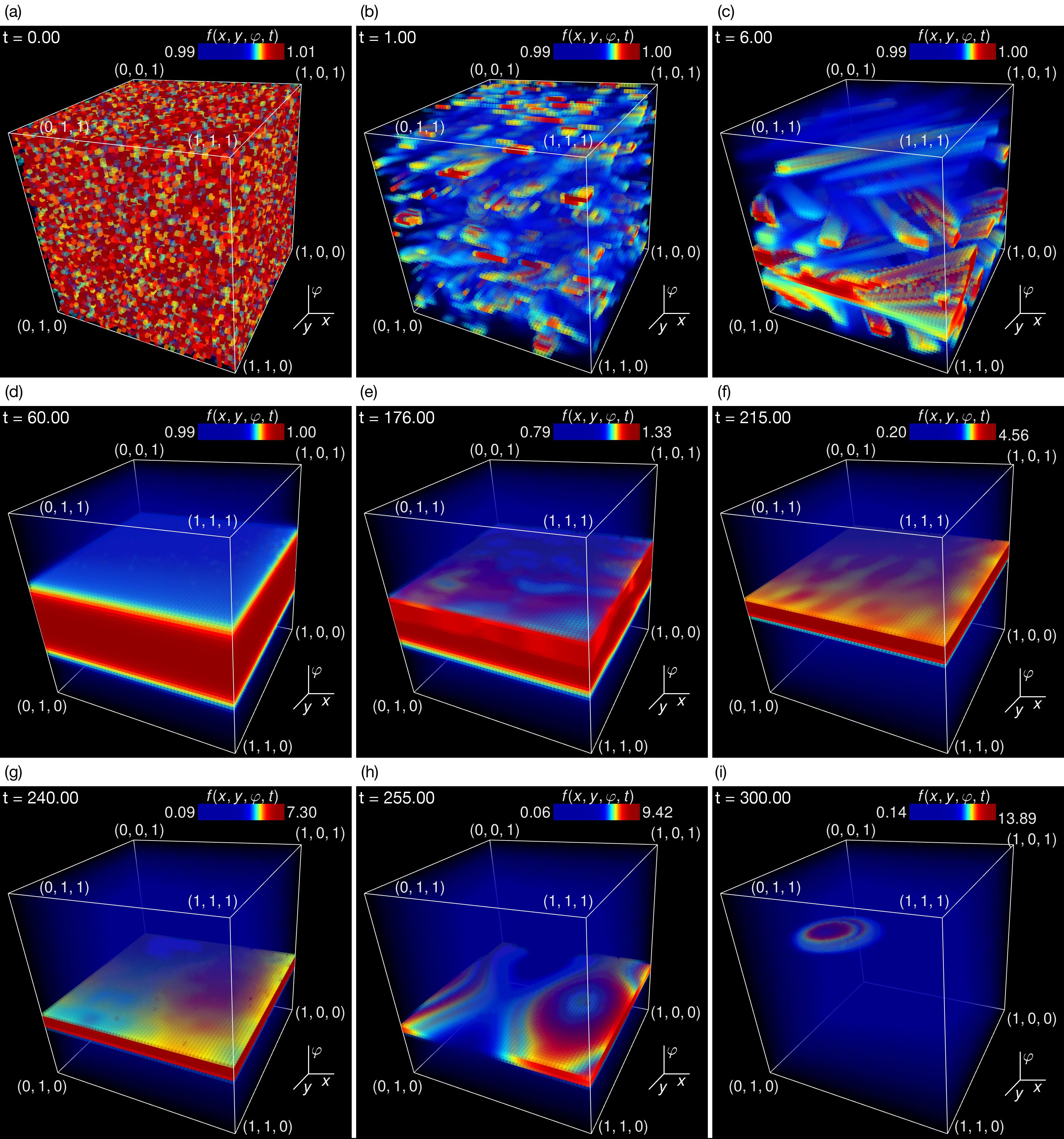}}
	\caption
	{
		\label{fig:complete_pde}
		Solution for the 3+1-dimensional one-particle density function from the continuum limit \eqref{eq:appendix:one-particle probability density function_transformed} in the regime of localized chimera with parameters $\sigma=4.0, \rho=0.3, \alpha=1.54, D_\varphi=0.01$. Color code corresponds to the density value, rescaled at each time point. Different snapshots represent the solution of the system at characteristic phases: (a) initial density values are assigned a constant value plus some perturbation; (b) most of the mass gathers in patches of high density; (c) the patches merge in tubes, which in turn begin to aggregate; (d) most of the mass gathers in a layer that is uniform in $(x,y)$; (e) the layer starts to shrink and wavy structures appear in front of it; (f) the layer compresses considerably and the waves diminish; (g) the layer becomes inhomogeneous in $(x,y)$; (h) the elongated structure of high density emerges; (i) the final high-density cluster of ellipsoidal shape forms and moves along the helical trajectory.
	}
\end{figure*}

This equation is still not closed with regard to $f^N$. To deal with the noise term, we define the new noise term as
\begin{equation*}
\xi(\vec{s},t) = -\frac{1}{N}\sum_{i=1}^N \nabla\cdot[\vec{\eta}_i \delta_i^3(\vec{s},t)],
\end{equation*}
which has the following correlation function
\begin{align*}
\bigl< \xi(\vec{s},t),\xi(\vec{s'},t') \bigr> &= \\
\frac{1}{N^2}\delta(t-t')&\sum_{i=1}^N\nabla_{\vec{s}}\cdot\nabla_{\vec{s'}}[\delta_i^3(\vec{s},t)\delta_i^3(\vec{s'},t)].
\end{align*}
Exploiting the property of the Dirac delta function $\delta^3(\vec{s}-\vec{s}_i,t)\delta^3(\vec{s'}-\vec{s}_i,t) = \delta^3(\vec{s}-\vec{s'})\delta(\vec{s}-\vec{s}_i)$, we obtain
\begin{equation*}
\bigl< \xi(\vec{s},t)\xi(\vec{s'},t') \bigr> = \frac{1}{N}\delta(t-t')\nabla_{\vec{s}}\cdot\nabla_{\vec{s'}}[f^N(\vec{s},t)\delta^3(\vec{s}-\vec{s'})].
\end{equation*}
Now we redefine the white Gaussian noise by introducing a global noise field
\begin{equation*}
\xi'(\vec{s},t) = \nabla\cdot\left[\vec{\eta}(\vec{s},t) \sqrt{\frac{1}{N}f^N(\vec{s},t)}\right].
\end{equation*}
Since the both noises $\xi(\vec{s},t)$ and $\xi'(\vec{s},t)$ have the same correlation functions, they are statistically identical. Therefore, we rewrite \eqref{eq:deriv_3} with the new noise field and obtain the closed-form expression for the microscopic phase space density function
\begin{equation}
\begin{aligned}
\label{eq:klimontovich}
&\frac{\partial f^N(\vec{s},t)}{\partial t} = -\nabla\cdot[f^N(\vec{s},t)\vec{u}] \\
- \nabla\cdot&\left[ \vec{\eta}(\vec{s},t) \sqrt{\frac{2D_\varphi f^N(\vec{s},t)}{N}} \right] + D_\varphi\Delta f^N(\vec{s},t),
\end{aligned}
\end{equation}
which is of the Klimontovich type \cite{nicholson:wiley} from plasma theory. 
However, the structure of this equation is different because of the non-linear alignment coming from the third component of $\vec{u}$, which is dependent on $\varphi$ itself.

The microscopic density \eqref{eq:klimontovich} contains the information about all the particles. We are interested in obtaining the evolution of a one-particle density function which depends only of $\vec{s},t$. Thus, we consider the ensemble averaged quantity $f(\vec{s},t)$ \cite{archer:jpa} obtained from taking an ensemble average of $f^N(\vec{s},t)$. In order to integrate all the particle dependencies out, we make the molecular chaos assumption and obtain the mean-field limit $N\rightarrow\infty$ \cite{spohn:springer}, where the term that scales as $1 / \sqrt{N}$ vanishes. Eventually, we have
\begin{align*}
\frac{\partial f(\vec{s},t)}{\partial t} &= -\nabla\cdot[f(\vec{s},t)\vec{w}] + D_\varphi\Delta f(\vec{s},t),
\end{align*}
where $\vec{w} = \begin{pmatrix} \cos\varphi \\ \sin\varphi \\ \sigma\cfrac{\iint\limits_{B_\rho}\int\limits_0^{2\pi} f^N\sin(\varphi' - \varphi - \alpha) \mathrm{d}x'\mathrm{d}y'\mathrm{d}\varphi'}{\iint\limits_{B_\rho}\int\limits_{0}^{2\pi} f^N \mathrm{d}x'\mathrm{d}y'\mathrm{d}\varphi'} \end{pmatrix}$.
This equation can be regarded as the Vlasov type equation \cite{nicholson:wiley}, or as the Fokker-Planck equation \cite{risken}. However, the same remark is applied here: the structure of this equation is different because of the non-linear alignment coming from the third component of $\vec{w}$, thus, we can not speak of the complete equivalence of equation types.
Note that at the beginning we have also implied that the noise acts only on the direction of motion $\varphi$, which suggests that $B_t^x = B_t^y \equiv 0$. Therefore, the Laplace operator only acts with respect to $\varphi$. Thus, if we restore the original variables, i.e. $x, y, \varphi$, we finally proceed to the following one-particle density function with a non-local coupling term:
\begin{equation}
\label{eq:appendix:one-particle probability density function}
\begin{aligned}
&\frac{\partial}{\partial t} f = - \frac{\partial}{\partial x}(f\cos\varphi) - \frac{\partial}{\partial y}(f\sin\varphi) \\
&- \frac{\partial}{\partial \varphi}\left(f \frac{\sigma}{\left\vert B_\rho \right\vert} \iint\limits_{B_\rho}\int\limits_0^{2\pi} f\sin(\varphi' - \varphi - \alpha) \mathrm{d}x'\mathrm{d}y'\mathrm{d}\varphi'\right) \\
&+ D_\varphi \frac{\partial^2}{\partial\varphi^2}f
\end{aligned}
\end{equation}
with the normalization term
\begin{align*}
\lvert B_\rho \rvert = \iint\limits_{B_\rho}\int\limits_{0}^{2\pi} f \mathrm{d}x'\mathrm{d}y'\mathrm{d}\varphi'.
\end{align*}

\subsection*{Complete solution}

One-particle probability density function \eqref{eq:appendix:one-particle probability density function} is defined in a cube with periodic boundaries of size $[0;1)^2\times[0;2\pi)$ for $x,y,\varphi$ coordinates, respectively. In order to have the same discretization level in all dimensions, we introduce the change of parameters
\begin{align*}
\tilde{\varphi} &= \frac{1}{2\pi}\varphi, \\
\tilde{f}(x,y,\tilde{\varphi},t) &= f(x,y,\varphi,t),
\end{align*}
so that the new density function $\tilde{f}$ is defined in a cube of size $[0;1)^3$. Taking the transformation into account, the Fokker-Planck equation \eqref{eq:appendix:one-particle probability density function} for the function $\tilde{f} = \tilde{f}(x,y,\tilde{\varphi},t)$ is transformed as follows
\begin{equation}
\label{eq:appendix:one-particle probability density function_transformed}
\begin{aligned}
\frac{\partial}{\partial t} \tilde{f} &= - \frac{\partial}{\partial x}[\tilde{f}\cos(2\pi\tilde\varphi)] - \frac{\partial}{\partial y}[\tilde{f}\sin(2\pi\tilde\varphi)] \\
- \frac{\partial}{\partial \tilde\varphi}&\left[ \frac{\tilde{f}\sigma}{2\pi \left\vert \tilde{B}_\rho \right\vert} \iint\limits_{B_\rho}\int\limits_0^{1} \tilde{f}\sin(2\pi\tilde\varphi' - 2\pi\tilde\varphi - \alpha) \mathrm{d}x'\mathrm{d}y'\mathrm{d}\tilde\varphi'\right] \\
&+ \frac{D_\varphi}{4\pi^2} \frac{\partial^2}{\partial\tilde\varphi^2}\tilde{f}
\end{aligned}	
\end{equation}
with the modified normalization term
\begin{align*}
\lvert \tilde{B}_\rho \rvert = \iint\limits_{B_\rho}\int\limits_{0}^{1} \tilde{f} \mathrm{d}x'\mathrm{d}y'\mathrm{d}\tilde\varphi'.
\end{align*}

In the main text, the solution of \eqref{eq:appendix:one-particle probability density function_transformed} is demonstrated in the form of a sequence of projections in the $(x,y)$-space.
In this section, we demonstrate the complete $3+1$-dimensional solution of the partial differential equation (cf. Fig.~\ref{fig:complete_pde}; see video S4 for the system behavior under \cite{supp_mat,bcs_youtube_channel}). In the case of a localized chimera scenario, which is of particular interest, the system transverses a series of ubiquitous stages until convergence. 
At the beginning, each grid point is initialized with a constant plus small amount of noise; then, the whole system is rescaled so that $\int\limits_0^1\int\limits_0^1\int\limits_0^1 \tilde{f}(x,y,\tilde\varphi,t) \mathrm{d}x\mathrm{d}y\mathrm{d}\tilde\varphi = 1$ holds (Fig.~\ref{fig:complete_pde}(a)). As the fluid starts to evolve, small patches of higher density appear (Fig.~\ref{fig:complete_pde}(b)). Those patches soon merge into tubes of high density which also gradually interflow (Fig.~\ref{fig:complete_pde}(c)) and later form a thick layer, which is homogeneous in $(x,y)$ (Fig.~\ref{fig:complete_pde}(d)). That layer slowly narrows and at some point the fluid in front of it starts to fluctuate (Fig.~\ref{fig:complete_pde}(e)) and the layer shrinks rapidly (Fig.~\ref{fig:complete_pde}(f)). Afterwards, the layer becomes inhomogeneous in $(x,y)$ (Fig.~\ref{fig:complete_pde}(g)) and disintegrates into an elongated form of high density (Fig.~\ref{fig:complete_pde}(h)). Eventually, that form disintegrates further and the high-density compact ellipsoidal shape appears and moves along a helical trajectory (Fig.~\ref{fig:complete_pde}(i)) through the environment of low density. By the comparison with the finite-size particle model, it resembles a group of particles that gather compactly together and travel in the same direction through the environment of chaotic uniformly distributed particles. Note that the density everywhere else in the system is not zero.

In the case of a non-localized chimera scenario, we observe a similar initial evolution until the high density layer emerges. But that layer does not disintegrate and is a final state for the system. By the comparison with the finite-size particle model, it means that there is a group of particles that synchronize in the direction of motion but are unable to travel locally together and, thus, remain scattered. Similarly to the previous case, the density everywhere else outside of that layer is not zero.

It should be mentioned that the direction $\tilde\varphi$ of maximum instantaneous density is not a unique value but rather a small range of values. It is so because there is diffusion included $D_\varphi=0.01$ and because the numerical scheme is itself dissipative and adds small amount of numerical diffusion.

\section*{Appendix D: Remarks to the videos}

The videos described in this section can be found under \cite{supp_mat,bcs_youtube_channel}. All videos, except for Video S5, were produced the following way. First, sequences of .png images were rendered by a C++ program through OpenGL. Then, the videos were produced from those sequences using the 'Blender' computer graphics software.  Video S5 was rendered in Matlab using VideoWriter object.

The videos illustrate the dynamics of self-propelled chimeras discussed throughout the paper. The considered domain for the finite-size particle model \eqref{eq:spc_sys} is a unit square $[0,1]\times[0,1]$ with periodic boundary conditions. The considered domain for the density function from the continuum limit \eqref{eq:appendix:one-particle probability density function_transformed} is $[0,1]\times[0,1]\times[0,1]$ with periodic boundary conditions. A color bar represents the absolute value of the averaged angular velocity $\left|\left<\dot{\varphi}\right>\right|$, subject to binary thresholding operation, for Videos S1-S3 and the density value $\tilde{f}(x,y,\tilde\varphi,t)$ for Videos S4-S5. All of the videos start with random initial conditions.

\subsection*{Video S1}

The dynamics of a system exhibiting a localized self-propelled chimera behavior. This solution has been obtained by integrating equation \eqref{eq:spc_sys} with the following parameters: $\sigma = 1.0$, $\rho = 0.3$, $\alpha = 1.54$, $N = 1000$. The output video format is H.264, the output file format is AVI, FFmpeg codec is H.264.

\subsection*{Video S2}

The dynamics of a system exhibiting a non-localized self-propelled chimera behavior. This solution has been obtained by integrating equation \eqref{eq:spc_sys} with the following parameters: $\sigma = 1.0$, $\rho = 0.03$, $\alpha = 1.5$, $N = 1000$. The output video format is H.264, the output file format is AVI, FFmpeg codec is H.264.

\subsection*{Video S3}

The dynamics of a system exhibiting a localized self-propelled chimera with increased coupling strength, which allows the rotation of a coherent group to be contained in one periodic box. This solution has been obtained by integrating equation \eqref{eq:spc_sys} with the following parameters: $\sigma = 4.0$, $\rho = 0.3$, $\alpha = 1.54$, $N = 1000$. The output video format is H.264, the output file format is AVI, FFmpeg codec is H.264.

\subsection*{Video S4}

The dynamics of a system exhibiting the equivalent to the localized self-propelled chimera in the continuum limit. The dynamics has been obtained by integrating equation \eqref{eq:appendix:one-particle probability density function_transformed} with the following parameters: $\sigma = 4.0, \rho = 0.3, \alpha = 1.54, D_\varphi = 0.01$. The spatial discretization is $60\times60\times60$. The output video format is MPEG, the output file format is MPEG-4.

\subsection*{Video S5}

The same dynamics as in Video S4 but projected into spatial coordinates $x,y$. The VideoWriter object has been set to MPEG-4 format.

\bibliographystyle{apsrev4-1}
\bibliography{self_propelled_chimeras}

\end{document}